\documentclass[letterpaper, 10 pt, conference]{ieeeconf}  

\IEEEoverridecommandlockouts                              
\usepackage{graphics} 
\usepackage{color} 
\usepackage{epsfig} 
\usepackage{mathptmx} 
\usepackage{times} 
\usepackage{amsmath} 
\usepackage{amssymb}  
\usepackage{enumerate}

\newtheorem{remark}{Remark}
\newtheorem{proposition}{Proposition}

\title{\LARGE \bf
MAP moving horizon state estimation with binary measurements
}

\author{Giorgio Battistelli, Luigi Chisci, Nicola Forti, Stefano Gherardini%
\thanks{The authors are with Dipartimento di Ingegneria dell'Informazione (DINFO), Universit\`a degli Studi di Firenze, via di Santa Marta 3, { \{giorgio.battistelli,luigi.chisci,nicola.forti,stefano.gherardini\}@unifi.it}. S. Gherardini is also
with Dipartimento di Fisica e Astronomia, Universit\`a degli Studi di Firenze, via G. Sansone 1, and Istituto Nazionale di Fisica Nucleare (INFN), Sezione di Firenze.}
}

\begin{document}

\maketitle
\thispagestyle{empty}
\pagestyle{empty}

\begin{abstract}
The paper addresses state estimation for discrete-time systems with binary (threshold) measurements by following a \textit{Maximum A posteriori Probability} (MAP) approach and exploiting a \textit{Moving Horizon} (MH) approximation of the MAP cost-function. It is shown that, for a linear system and noise distributions with log-concave probability density function, the proposed MH-MAP state estimator involves the solution, at each sampling interval, of a convex optimization problem. Application of the MH-MAP estimator to dynamic estimation of a diffusion field given pointwise-in-time-and-space binary measurements of the field is also illustrated and, finally, simulation results relative to this application are shown to demonstrate the effectiveness of the proposed approach.
\end{abstract}

\section{Introduction}
Binary sensors, whose output just indicates whether the noisy measurement of the sensed variable (analog measurement) exceeds or not a given threshold, are frequently employed in monitoring and control applications \cite{Wang1}-\cite{Capponi}.
The idea is that by a multitude of low-cost and low-resolution sensing devices it is possible to achieve the same estimation accuracy
that a few (possibly a single one) expensive high-resolution sensors could provide, with significant practical benefits in terms of ease of sensor deployment and minimization of communication requirements.
The fact that a binary (threshold) measurement just conveys a minimal amount (i.e. a single bit) of information, while implying communication bandwidth savings and consequently greater energy efficiency, makes of paramount importance to fully exploit the little available information by means of smart estimation algorithms.
In this respect, some work has recently addressed system identification \cite{Wang1}-\cite{Wang2}, parameter \cite{Ristic}-\cite{Ribeiro2} or state estimation \cite{Irr-sampling}-\cite{Capponi} with binary measurements by following either
a deterministic \cite{Wang1}-\cite{Wang2}, \cite{Irr-sampling}-\cite{Bai} or a probabilistic \cite{Ristic}-\cite{Ribeiro2}, \cite{Aslam}-\cite{Capponi} approach.

In a deterministic context, the available information is essentially concentrated at the sampling instants in which some binary measurement signal has switched value \cite{Irr-sampling,Gherardini}.
As shown in \cite{Gherardini}, some additional information can be exploited in the other (non switching) sampling instants
by penalizing values of the estimated quantity such that the corresponding predicted measurement is on the opposite side, with respect to a binary sensor reading, far away from the threshold.
Nevertheless, it is clear that there is no or very little information available for estimation purposes whenever no or very few binary sensor switchings occur.
Hence, a possible way to achieve high estimation accuracy is to have many binary sensors measuring the same variable with different thresholds as this would clearly increase the number of switchings, actually emulating, when the number of sensors tends to infinity, the availability of a single continuous-valued (analog) measurement.

Conversely, following a probabilistic approach, binary sensor readings could be exploited to infer information about the probability
distribution of the variable of interest.
To clarify this point, let us assume that a very large number of binary sensors of the same type (i.e. measuring the same variable with the same threshold) be available and the distribution of their measurement noise (e.g. Gaussian with zero mean and given standard deviation) be known.
Then, thanks to the numerosity of measurements, the relative frequency of $1$ (or $0$) values occurring in the sensor readings could be considered as a reasonable estimate of the probability that the sensed variable is above (or below) the threshold and this, in turn, exploiting the knowledge of the measurement noise distribution allows to extract information about the location of the value of the sensed variable with respect to the threshold.
If, for example, it is found that the binary measurement is equal to $1$ for $70 \%$ of the sensors and Gaussian measurement noise is
hypothesized, it turns out that the expected measurement of the sensed variable is above the threshold of an amount equal to
$0.525$ times the standard deviation of the measurement noise.
Notice that if the sensors are noiseless, they all provide either $0$ or $1$ output  and, paradoxically, in this case minimal information, i.e. that the sensed variable takes values in a semi-infinite interval (either below or above the threshold), is extracted from the set of binary measurements.
The above arguments suggest that, adopting a probabilistic approach to estimation using binary measurements, the presence of measurement noise can be a helpful source of information.
In other words, it can be said that noise-aided procedures can be devised for estimation with binary measurements by exploiting the fact that the measurement noise randomly shifts the analog measurement thus making possible to infer statistical information on the sensed variable.

Relying on the above stated \textit{noise-aided} paradigm, this paper presents a novel approach to recursive estimation of the state of a discrete-time dynamical system given binary measurements.
The proposed approach is based on a \textit{moving-horizon} (MH) approximation of  the \textit{Maximum A-posteriori Probability} (MAP) estimation and extends previous work \cite{Ristic}-\cite{Wong} concerning parameter estimation to recursive state estimation. A further contribution of the paper is to show that for a linear system the optimization problem arising from the MH-MAP formulation
turns out to be convex and, hence, practically feasible for real-time implementation.

The rest of the paper is organized as follows.
Section II introduces the MAP problem formulation of state estimation with binary measurements.
Section III presents a MH approximation of MAP estimation, referred to as MH-MAP algorithm, and analyzes the properties of the resulting optimization problem.
Section IV discusses a possible application of the proposed approach to the dynamic estimation of a diffusion field from binary pointwise-in-space-and-time field measurements.
Section V presents simulation results relative to the dynamic field estimation case-study.
Finally, section VI concludes the paper with perspectives for future work.

\section{Maximum a posteriori state estimation with binary sensors}

The following notation will be used throughout the paper: $col(\cdot)$ denotes the matrix obtained by stacking its arguments one on top of the other; $diag(m^{1},\ldots,m^{q})$ denotes the diagonal matrix with diagonal entries $m^{1},\ldots,m^{q}$; $0_{n}$, $1_{n}$ indicate the $n-$dimensional vectors, respectively, with all zero and unit entries.

Let us consider the problem of recursively estimating the state of the discrete-time nonlinear dynamical system
\begin{eqnarray}
x_{t+1} & = & f(x_{t},u_{t})+ w_{t}\label{1} \\
z_{t}^{i}     & = & h^{i}(x_{t})+ v_{t}^{i},\hspace{3mm}i=1,\ldots,l \label{2}
\end{eqnarray}
from a set of measurements provided by binary sensors
\begin{equation} \label{3}
\begin{array}{rclcl}
y_{t}^{i} & = & g^i(z_{t}^{i}) & = &  \left\{
\begin{array}{ll} 1, & \mbox{if }  z_{t}^{i} \geq \tau^{i} \\
                            0,   & \mbox{if }  z_{t}^{i}  < \tau^{i}
\end{array}
\right.
\end{array}
\end{equation}
where $x_t \in \mathbb{R}^n$ is the state to be estimated, $u_t \in \mathbb{R}^m$ is a known input, and $\tau^i$ is the threshold of the $i-$th binary sensor. For the sake of simplicity, we define $z_t = col \left( z_t^i \right)_{i=1}^l \in \mathbb{R}^l$ and $y_t = col \left( y_t^i \right)_{i=1}^l \in \mathbb{R}^l$ . The vector $w_{t}\in \mathbb{R}^n$ is an additive disturbance affecting the system dynamics which accounts for uncertainties in the mathematical model, while $v_t = col \left( v_t^i \right)_{i=1}^l\in \mathbb{R}^l$ is the measurement noise vector.

Let $\mathcal{N}(\mu,\Sigma)$ denote as usual the normal distribution with mean $\mu$ and variance $\Sigma$.
The statistical behaviour of the system is characterized by
\begin{equation}
x_{0}\sim\mathcal{N}(\overline{x}_{0},P^{-1}),\hspace{2mm}w_{t}\sim\mathcal{N}(0,Q^{-1}),\hspace{2mm}v_{t}\sim\mathcal{N}(0,R)
\label{prob}
\end{equation}
where: $R=diag(r^{1},\ldots,r^{p})$; $\mathbb{E}[w_{j}w_{k}']=0$ and $\mathbb{E}[v_{j}v_{k}']=0$ if $j\neq k$; $\mathbb{E}[w_{j}v_{k}']=0$, $\mathbb{E}[w_{j}x_{0}']=0$, $\mathbb{E}[v_{j}x_{0}']=0$ for any $j,k$.
Notice from (\ref{2})-(\ref{3}) that sensor $i$ produces a binary measurements $y^{i}_{t}\in\{0,1\}$ depending on whether the noisy system output $z^{i}_{t}$ is below or above the threshold $\tau^{i}$.

According to the available probabilistic description (\ref{prob}), the problem of estimating the state of system  (\ref{1}) under the binary measurement model  (\ref{2})-(\ref{3}) is formulated hereafter in a Bayesian framework by resorting to a maximum a posteriori probability (MAP) criterion.
In the remainder of this section, as a preliminary step, the {\em full-information} MAP state estimation problem is formulated.

To this end, notice that each binary measurement $y^{i}_t$ provides intrinsically  relevant information on the state $x_t$ which can be taken into account by means of the a posteriori probabilities $p(y^{i}_{t}|x_{t})$. In particular, the binary measurement $y^{i}_{t}$ is a Bernoulli random variable such that, for any binary sensor $i$ and any time instant $t$, the a posteriori probability $p(y^{i}_{t}|x_{t})$ is given by
\begin{equation}
p(y^{i}_{t}|x_{t})~=~p(y^{i}_{t}=1|x_{t})^{y^{i}_{t}} ~p(y^{i}_{t}=0|x_{t})^{1-y^{i}_{t}}
\end{equation}
where
\begin{equation}\label{8}
p(y^{i}_{t}=1|x_{t}) = F^i(\tau^{i}-h^{i}(x_{t}))
\end{equation}
and $p(y^{i}_{t}=0|x_{t})=1-p(y^{i}_{t}=1|x_{t})\triangleq \Phi^i(\tau^{i}-h^{i}(x_{t})) $. The function $F^i(\tau^{i}-h^{i}(x_{t}))$ is the complementary cumulative distribution function (CDF) of the random variable $\tau^{i}-h^{i}(x_{t})$. Since $v^{i}_{t}\sim\mathcal{N}(0,r^{i})$, the conditional probability $p(y^{i}_{t}=1|x_{t}) = F^i(\tau^{i}-h^{i}(x_{t})) $ can be written in terms of the Q-function as follows
\begin{equation}
F^i(\tau^{i}-h^{i}(x_{t}))  = \frac{1}{\sqrt{2\pi r^{i}}}\int_{\tau^{i}-h^{i}(x_{t})}^{\infty}e^{-\frac{u^{2}}{2r^{i}}}du= Q \left(\frac{\tau^{i}-h^{i}(x_{t})}{\sqrt{r^{i}}}\right) \, .
\end{equation}

Let us now denote by $Y_{t}={\rm col} (y_0,\ldots,y_{t} )$ the vector of all binary measurements collected up to time $t$ and by  $X_{t}\triangleq {\rm col} (x_{0},\ldots,x_{t} )$ the vector of the state trajectory.
Further, let us denote by $\hat{X}_{t|t} \triangleq {\rm col} ( \hat{x}_{0|t},\ldots,\hat{x}_{t|t} )$ the estimates of $X_{t}$ to be made at any stage $t$. Then,
at each time instant $t$, given the a posteriori probability $p(X_{N}|Y_{N})$, the estimate of the state trajectory can be obtained by solving the following MAP estimation problem:
\begin{equation}
\hat{X}_{t|t}  = \text{arg}\max_{X_{t}}p(X_{t}|Y_{t})
=\text{arg} \min_{X_{t}}  - \ln p(X_{t}|Y_{t}) \label{12}.
\end{equation}
From the Bayes rule
\begin{equation}
p(X_{t}|Y_{t}) ~\propto ~p(Y_{t}|X_{t})~p(X_{t}),
\end{equation}
where
\begin{equation}
\begin{split}
p(X_{t})&=\prod_{k=0}^{t-1}p(x_{t-k}|x_{t-k-1},\ldots,x_{0})~p(x_{0}) \\
&=\prod_{k=0}^{t-1}p(x_{t-k}|x_{t-k-1})~p(x_{0}).
\end{split}
\end{equation}
Notice that in the latter equation we have considered the Markov property for the dynamical system state. As $x_{0}$ and $w_{t}$ are normally distributed vectors, we have
\begin{eqnarray}
&&p(x_0) \propto e^{-\frac{1}{2}\|x_{0}-\overline{x}_0 \|^{2}_{P}}\label{15} \\
&&p(x_{k}|x_{k-1}) \propto e^{-\frac{1}{2}\|x_{k+1}-f(x_{k},u_{k})\|^{2}_{Q}}\label{16} .
\end{eqnarray}

Moreover, the likelihood function $p(Y_{t}|X_{t})$ of the binary measurement vector $Y_{t}$ can be written as
\begin{equation}
\begin{split}
p(Y_{t}|X_{t})&=\prod_{k=0}^{t}p(y_{k}|x_{k})=\prod_{k=0}^{t}~\prod_{i=1}^{l}p(y_{k}^{i}|x_{k}) \\
&=\prod_{k=0}^{t}~\prod_{i=1}^{l}F^i(\tau^{i}-h^{i}(x_{k}))^{y_{k}^{i}} ~ \Phi^i(\tau^{i}-h^{i}(x_{k}))^{1-y_{k}^{i}}
\end{split}
\end{equation}
where in the latter equality we have exploited the statistical independence of the binary sensors. Accordingly, the log-likelihood is
\begin{equation}
\begin{split}
\ln p(Y_{t}|X_{t})=&\sum_{k=0}^{t}~\sum_{i=1}^{l}\left[y_{k}^{i} \, \ln F^i (\tau^{i}-h^{i}(x_{k}))\right. \\
&\left.+(1-y_{k}^{i}) \, \ln \Phi^i (\tau^{i}-h^{i}(x_{k}))\right],
\end{split}
\end{equation}
and the cost function $- \ln p(X_{t}|Y_{t})  =-\ln p(Y_{t}|X_{t})-\ln p(X_{t})$ to be minimized in the MAP estimation problem (\ref{12}) turns out to be, up to additive constant terms,
\begin{equation}\label{18}
\begin{split}
J_{t} (X_t) &=\|x_{0}-\overline{x}_{0}\|^{2}_{P}+\sum_{k=0}^{t}\|x_{k+1}-f(x_{k},u_{k})\|^{2}_{Q}+ \\
&\hspace{-1cm} -\sum_{k=0}^{t}~\sum_{i=1}^{l}\left[y_{k}^{i} \ln F^i(\tau^{i}-h^{i}(x_{k}))+(1-y_{k}^{i}) \ln \Phi^i(\tau^{i}-h^{i}(x_{k}))\right] \, .
\end{split}
\end{equation}

Unfortunately, a closed-form expression for the global minimum of (\ref{18}) does not exist and, hence, the optimal MAP estimate $\hat X_{t|t}$ has to be determined by resorting to some numerical
optimization routine. With this respect, the main drawback is that the number of optimization variables grows linearly with time, since the vector $X_t$ has size $(t+1) \, n$. As a consequence,
as $t$ grows the solution of the full information MAP state estimation problem (\ref{12}) becomes eventually unfeasible, and some approximation has to be introduced.

\section{Moving-horizon approximation}

In this section, an approximate solution to the  MAP state estimation problem is proposed by resorting to the MHE approach
\cite{Morari}-\cite{Delgado}.
Accordingly, by defining a sliding window $\mathfrak W_t = \{t-N, t-N+1, \ldots, t\}$,  the goal is to find an estimate of the partial state trajectory $X_{t-N:t} \triangleq {\rm col} ( x_{t-N},\ldots,x_{t} ) $
by using the information available in $\mathfrak W_t$. Then, in place of the full information cost $J_t (X_t)$, at each time instant $t$ the minimization of the following {\em moving-horizon cost} is addressed:
 \begin{equation}\label{MH}
\begin{split}
J_{t}^{\rm MH} (X_{t-N:t}) &= \Gamma_{t-N} (x_{t-N}) + \sum_{k=t-N}^{t}\|x_{k+1}-f(x_{k},u_{k})\|^{2}_{Q}+ \\
&\hspace{-2cm} -\sum_{k=t-N}^{t}~\sum_{i=1}^{l}\left[y_{k}^{i} \ln F^i(\tau^{i}-h^{i}(x_{k}))+(1-y_{k}^{i}) \ln \Phi^i(\tau^{i}-h^{i}(x_{k}))\right]
\end{split}
\end{equation}
where the non-negative initial penalty function $\Gamma_{t-N} (x_{t-N}) $, known in the MHE literature as {\em arrival cost} \cite{Rao2,CDC},
is introduced so as to summarize the past data $y_0, \ldots, y_{t-N-1}$ not explicitly accounted for in the objective function.

As a matter of fact, the form of the arrival cost plays an important role in the behavior and performance of the overall estimation scheme.
While in principle $\Gamma_{t-N} (x_{t-N})$ could be chosen so that minimization of (\ref{MH}) yields the same estimate that would be obtained by minimizing (\ref{18}),
an algebraic expression for such a true arrival cost seldom exists, even when the sensors provide continuous (non-binary) measurements \cite{Rao2}.
Hence, some approximation must be used. With this respect, a common choice \cite{NLMHE,CDC}, also followed in the present work,
consists of assigning to the arrival cost a fixed structure penalizing the distance of the state $x_{t-N}$ at the beginning of the sliding window
from some prediction $\bar x_{t-N} $ computed at the previous time instant, thus making the estimation scheme recursive.
A natural choice is then a quadratic arrival cost of the form
\begin{equation}\label{eq:arrival}
\Gamma_{t-N} (x_{t-N}) = \|x_{t-N}-\bar {x}_{t-N}\|^{2}_{\Psi} \, ,
\end{equation}
which, from the Bayesian point of view, corresponds to approximating the PDF of the state $x_{t-N}$ conditioned to all the measurements collected up
to time $t-1$ with a Gaussian having mean $\bar {x}_{t-N}$ and covariance $\Psi^{-1}$. As for the choice of the weight matrix $\Psi$,
in the case of continuous measurements it has been shown that stability of the estimation error dynamics can be ensured provided that $\Psi$
is not too large (so as to avoid an overconfidence on the available estimates) \cite{NLMHE,CDC}. Recently \cite{Gherardini}, similar results have been proven to hold also in the case of
binary sensors in a deterministic context. In practice, $\Psi$ can be seen as a design parameter which has to be tuned by pursuing a suitable tradeoff between such
stability considerations and the necessity of not neglecting the already available information (since in the limit for $\Psi$ going to zero the approach becomes
a finite memory one).

Summing up,  at any stage $t=N,N+1,\ldots$, the following problem has to be solved.  \vspace{.3cm}

\textbf{Problem $E_{t}$:} Given the prediction $\bar{x}_{t-N}$, the input sequence $\{ u_{t-N}, \ldots, u_{t-1} \}$, the measurement sequences $\{ y^i_{t-N}, \ldots, y^i_t , \, i = 1, \ldots, l \}$,
find the optimal estimates $\hat{x}_{t-N|t},\ldots,\hat{x}_{t|t}$ that minimize the cost function (\ref{MH}) with arrival cost (\ref{eq:arrival}).
\vspace{.3cm}

Concerning the propagation of the estimation procedure from Problem $E_{t-1}$ to Problem $E_{t}$, the prediction $\bar{x}_{t-N}$ is set equal to the value of the estimate of $x_{t-N}$ made at time instant $t-1$, i.e., $\bar{x}_{t-N} = \hat{x}_{t-N|t-1}$. Clearly, the recursion is initialized with the a priori expected value $\bar x_0$ of the initial state vector.

In general, solving Problem $E_{t}$ entails the solution of a non-trivial optimization problem. However, when both equations (\ref{1}) and (\ref{2}) are linear, the resulting optimization problem turns out to be convex so that standard optimization routines can be used in order to find the global minimum. To see this, let us consider the following assumption.
\vspace{.3cm}

\begin{enumerate}[\bf {A}1]
\item The functions $f(\cdot)$ and $h^i(\cdot)$, $i=1,\ldots,l$, are linear, i.e., $f(x_t,u_t) = A x_t + B u_t$ and $ h^i(x_t) = C^i x_t$, $i=1,\ldots,l$, where $A$, $B$, $C^i$ are constant matrices of suitable dimensions.
\end{enumerate} \vspace{.3cm}

\begin{proposition}
If assumption A1 holds, the CDF $\Phi^i(\tau^{i}-C^{i}x_{t})$ and its complementary function $F^i(\tau^{i}-C^{i}x_{t})$ are log-concave. Hence,
the cost function (\ref{MH}) with arrival cost (\ref{eq:arrival}) is convex. \\
\mbox{   }\hfill $\square$
\end{proposition} \vspace{.3 cm}

\begin{remark}
Under assumption A1, the convexity of the cost function  (\ref{MH})  is guaranteed also in the more general case in which the statistical behaviour of the random variables $x_{0}$, $w_{t}$, $v_{t}$ is described by logarithmically concave distribution functions. Indeed, if a PDF is log-concave, also its cumulative distribution function is log-concave; hence the contribution related to the binary measurements in  (\ref{MH}) turns out to be convex.
\end{remark} \vspace{.3 cm}

In the next section we will focus on the case of a discrete-time linear system, in particular considering  a diffusion process governed by a \textit{partial differential equation} (PDE) and spatially discretized by means of the \textit{finite element method} (FEM).

\section{Dynamic field estimation}

In this section, we consider the problem of reconstructing a two-dimensional diffusion field, sampled with a network of binary sensors arbitrarily deployed over the spatial domain of interest $\Omega$. The diffusion process is governed by the following parabolic PDE:
\begin{equation}
\dfrac{\partial c}{\partial t} - \lambda \nabla^2 c  ~=~ 0 \,\,\,\,\, \mbox{in } \Omega
\label{PDE}
\end{equation}
which models various physical phenomena such as the spread of a pollutant in a fluid. In this case, $c(\xi,\eta,t)$ represents the space-time dependent substance concentration, $\lambda$ denotes the constant diffusivity of the medium, and $\nabla^2 = {\partial^2 } / {\partial \xi^2} + {\partial^2 } / {\partial \eta^2}$ is the Laplace operator, $(\xi,\eta) \in \Omega$ being the 2D spatial variables.
Furthermore, let us assume mixed boundary conditions, i.e. a non-homogeneous Dirichlet condition
\begin{equation}
c = \psi \,\,\,\,\,\,\, \mbox{ on } \partial \Omega_D,
\label{Dbc}
\end{equation}
which specifies a constant-in-time value of concentration on the boundary $\partial \Omega_D$,
and a homogeneous Neumann condition on $\partial \Omega_N = \partial \Omega \setminus \partial \Omega_D$, assumed impermeable to the contaminant, so that
\begin{equation}
{\partial c}/{\partial \upsilon} = 0 \,\,\,\,\,\,\, \mbox{ on } \partial \Omega_N,
\label{Nbc}
\end{equation}
where $\upsilon$ is the outward pointing unit normal vector of
$\partial \Omega_N$.

The objective is to estimate the values of the dynamic field of interest $c(\xi,\eta,t)$, given the binary measurements (\ref{3}). The PDE system (\ref{PDE})-(\ref{Nbc}) is simulated with a mesh of finite elements over $\Omega$  via the Finite Element (FE) approximation described in \cite{source}-\cite{ECC15}. Specifically, the domain $\Omega$ is subdivided into a suitable set of non overlapping regions, or elements, and a suitable set of basis functions $\phi_{j} (\xi , \eta)$,  $j=1,\ldots,n_\phi$ is defined on such elements. The choices of the basis functions $\phi_{j}$ and of the elements are key points of the FE method. In the specific case under investigation, the elements
are triangle in 2D and define a FE mesh with vertices $({\xi}_j, \eta_j) \in \Omega, j=1,\ldots,n_\phi $. Then each basis function $\phi_{j}$ is a piece-wise affine function which vanishes outside the FEs around $({\xi}_j, \eta_j)$ and such that $\phi_{j}({\xi}_j, \eta_j)=
\delta_{ij}$, $\delta_{ij}$ denoting the Kronecker delta. In order to account for the mixed boundary conditions, the basis functions are supposed to be ordered so that
the first $n$ correspond to vertices of the mesh which lie either in the interior of $\Omega$ or on $\partial \Omega_N$, while the last $n_\phi-n$ correspond to the vertices lying on $\partial \Omega_D$.

Accordingly, the unknown function $c(\xi \eta,t)$ is approximated as
\begin{equation}
c(\xi,\eta,t) \approx \sum_{j=1}^{n} \phi_{j}(\xi, \eta) \, c_j(t) + \sum_{j=n+1}^{n_\phi} \phi_{j}(\xi, \eta) \, \psi_j(t)
\label{EXPA}
\end{equation}
where $c_j(t)$ is the unknown expansion coefficient of the function $c(\xi \eta,t)$
relative to time $t$ and basis function $\phi_j(\xi,\eta)$, and $\psi_j$ is the known expansion coefficient of the
function $\psi(\xi \eta)$ relative to to the basis function $\phi_j(\xi,\eta)$. Notice that the second summation in
(\ref{EXPA}) is needed so as to impose the non-homogeneous Dirichlet condition (\ref{Dbc}) on the boundary $\partial \Omega_D$.

The PDE (\ref{PDE}) can be recast into the
following integral form:
\begin{equation}
\int_\Omega \frac{\partial c}{\partial t} \varphi \, d\xi d\eta  \, - \,
\lambda \int_\Omega ~\nabla^2 c~ \varphi \, d\xi d\eta  =0
\end{equation}
where $\varphi(\xi,\phi)$ is a generic space-dependent weight function.
By applying Green's identity,
one obtains:
\begin{equation}\label{eq:Green}
\int_\Omega \frac{\partial c}{\partial t} \varphi \, d\xi d\eta + \lambda
\int_\Omega \nabla^T c ~\nabla \varphi \, d\xi d\eta
- \lambda
\int_{\partial \Omega} \frac{\partial c}{\partial \upsilon} \varphi \,
d\xi d\eta
= 0 \, .
\end{equation}

By choosing the test function $\varphi$ equal to the selected basis functions and exploiting the approximation (\ref{EXPA}), the Galerkin
weighted residual method is applied and the following equation is obtained
\begin{eqnarray}
&& \sum_{i=1}^{n}  \int_\Omega \phi_i \phi_j \, d\xi d\eta \, \dot{c}_i(t)  + \lambda \sum_{i=1}^{n}  \int_\Omega \nabla^T \phi_i ~\nabla \phi_j \, d\xi d\eta \, c_i (t)  \nonumber \\
&& { } + \lambda  \sum_{i=n+1}^{n_\phi}   \int_\Omega \nabla^T \phi_i ~\nabla \phi_j \, d\xi d\eta \, \psi_i(t) = 0 \label{eq:Galerkin}
\end{eqnarray}
for $j = 1, \ldots, n$. Notice that in the latter equation the boundary integral of equation (\ref{eq:Green})
is omitted since it is equal to 0 thanks to the homogeneous Neumann condition (\ref{Nbc}) on $\partial \Omega_N$
and to the fact that, by construction, the basis functions $\phi_j$, $j = 1, \ldots, n$ vanish on $\partial \Omega_D$.
The interested reader is referred to \cite{Brenner96} for further details on the FEM theory, and in particular on how to convert the case of inhomogeneous boundary conditions to the homogeneous one.

By defining the state vector $x = {\rm col} (c_1 , \ldots , c_n) $ and the vector of boundary conditions $\gamma =  {\rm col} (\psi_{n+1}, \ldots, \psi_{n_\phi})$,
equation (\ref{eq:Galerkin}) can be written in the more compact form
\[
M \dot x (t) + S x(t) + S_D \gamma = 0
\]
where $S$ is the so-called stiffness matrix representing diffusion, $M$ is the mass matrix,  and $S_D$ captures the physical interconnections
among the vertices affected by boundary condition (\ref{Dbc}) and the remaining nodes of the mesh.

By applying for example the implicit Euler method, the latter equation can be discretized in time, thus obtaining the
the linear discrete-time model
\begin{equation}
x_{t+1} = A \, x_t + B \, u + w_t
\label{dt-sys}
\end{equation}
where
\begin{equation*}
\begin{aligned}
A &= \left[ I +\delta t ~M^{-1}S\right]^{-1} \\
B &= \left[ I +\delta t ~M^{-1}S\right]^{-1}M^{-1}\delta t \\
u & = - S_D ~\gamma ~
\end{aligned}
\label{eq:def}
\end{equation*}
$\delta t$ is the time integration interval, and $w_t$ is the process disturbance taking into account also the space-time discretization errors.

Notice that the linear system (\ref{dt-sys}) has dimension $n$ equal to the number of vertices of the mesh not lying on $\partial \Omega_D$.
The linear system (\ref{dt-sys}) is assumed to be monitored by a network of $l$ threshold sensors. Each sensor, before binary quantization is applied,
directly measure the pointwise-in-time-and-space concentration of the contaminant
in a point of the spatial domain $\Omega$. By exploiting (\ref{EXPA}), such a concentration can be written as a linear combination of the concentrations on the grid points,
so that the resulting output function takes the form
\begin{equation}
z_{t}^{i}      = C^{i} x_{t} + v_{t}^{i},\hspace{3mm}i=1,\ldots,l \label{z}
\end{equation}
and assumption A1 is fulfilled.

\section{Numerical results}

\begin{figure}[h!]
	\centering
	\includegraphics[scale=0.27]{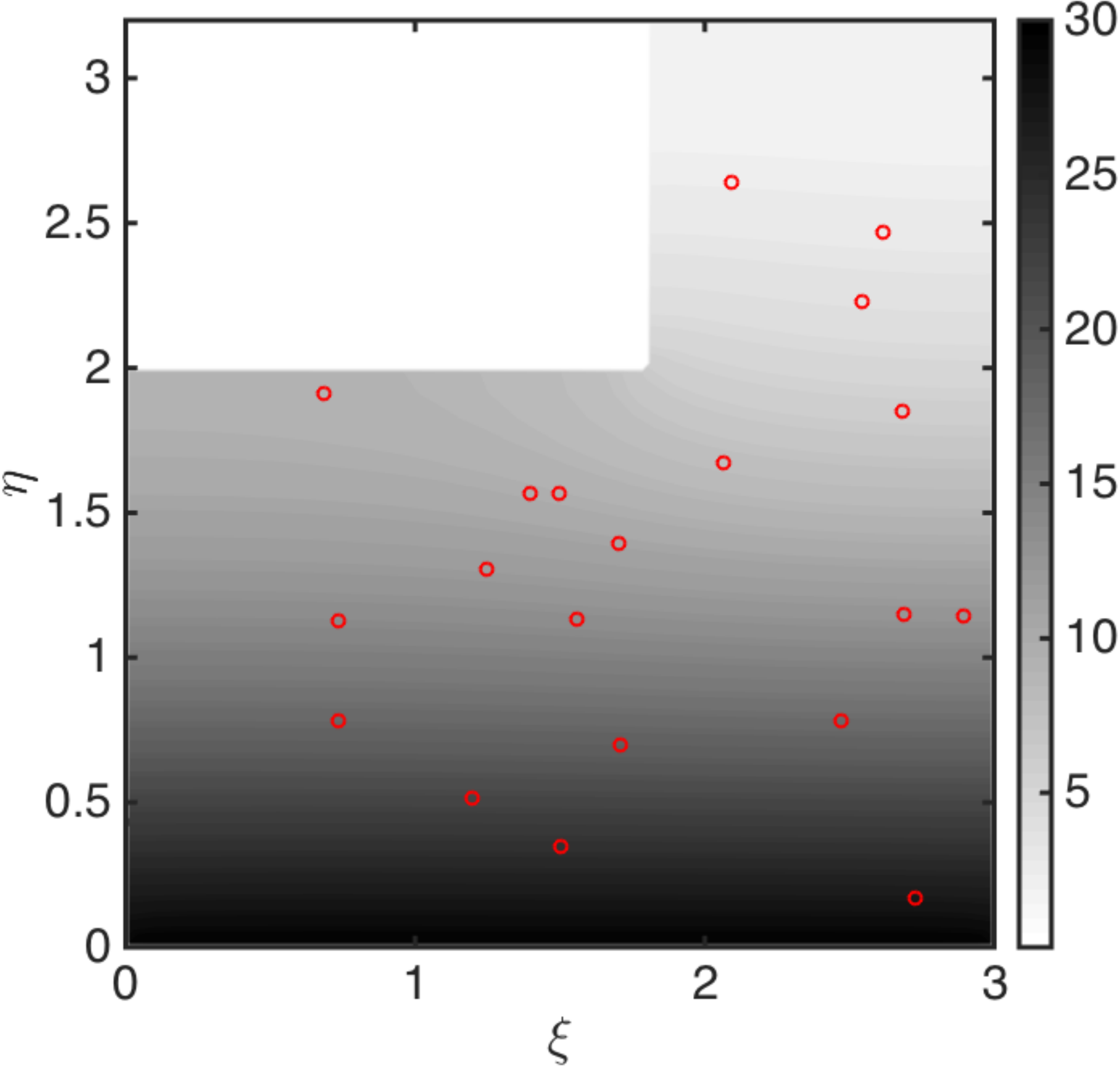}
	\caption{Concentration field at time $t = 100 ~ [s]$ monitored by a random network of 20 binary sensors (red $\circ$).}
	\label{fig:field200}
\end{figure}

\begin{figure}[t]
	\centering
	\includegraphics[scale=0.27]{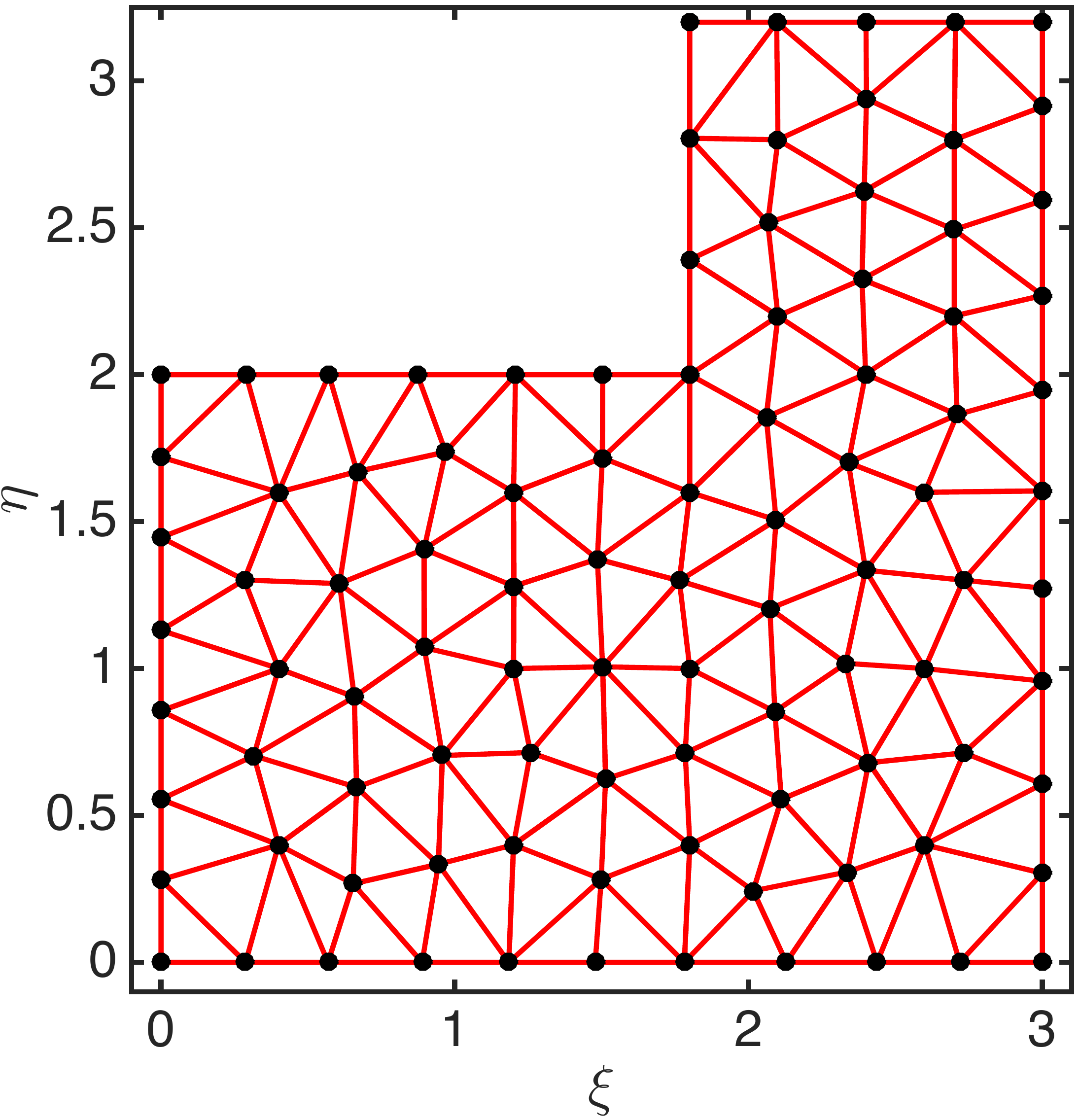}
	\caption{Mesh used by the MAP-MHE estimator (152 elements, 97 nodes).}
	\label{fig:mesh}
\end{figure}

In this section, we present the simulation results of the proposed approach applied to the problem of state estimation of spatially distributed processes, discussed in the previous section.
We consider the simulated system (\ref{dt-sys})-(\ref{z}) with $1695$ triangular elements, $915$ vertices, $\lambda = 0.01 ~ [m^2/s]$, fixed integration step length $\delta t = 1 ~ [s]$, $\gamma = 30 ~ [g/m^2]$, and
initial condition of the field vector $x_{0} = 0_n ~ [g/m^2]$. The field of interest is defined over a bounded 2D spatial domain $\Omega$ which covers an area of $7.44 ~ [m^2]$ (see Fig. \ref{fig:field200}), with boundary condition (\ref{Dbc}) on the bottom edge and no-flux condition (\ref{Nbc}) on the remaining portions of $\partial \Omega$.
Compared to the \textit{ground truth} simulator, the proposed MH-MAP estimator implements a coarser mesh (see Fig. \ref{fig:mesh}) of 97 nodes ($n = 89$, $d = 8$), and runs at a slower sample rate ($0.1 ~ [Hz]$), so that model uncertainty is taken into account. The initial condition of the estimated dynamic field is set to $\overline{x}_{0} = 5 ~ {1}_n ~ [g/m^2]$, the moving window has size $N = 5$, and the weight matrices in (\ref{prob}) are chosen as $\Psi = 10^3 ~ I_n$ and $Q = 10^{2} ~ I_n$.
The \textit{true} concentrations from (\ref{dt-sys}) are first corrupted with a Gaussian noise with variance $r^i$, then binary observations are obtained by applying a different threshold $\tau^i$ for each sensor $i$ of the network. Note that, in order to receive informative binary measurements, $\tau^i, ~ i = 1,...,l$ are generated as uniformly distributed random numbers in the interval $(0.05,29.95)$, being $(0,30)$ the range of nominal concentration values throughout each experiment. The duration of each simulation experiment is fixed to $1200 ~ [s]$ (120 samples).
\begin{figure}[t]
	\centering
	\includegraphics[width=\columnwidth]{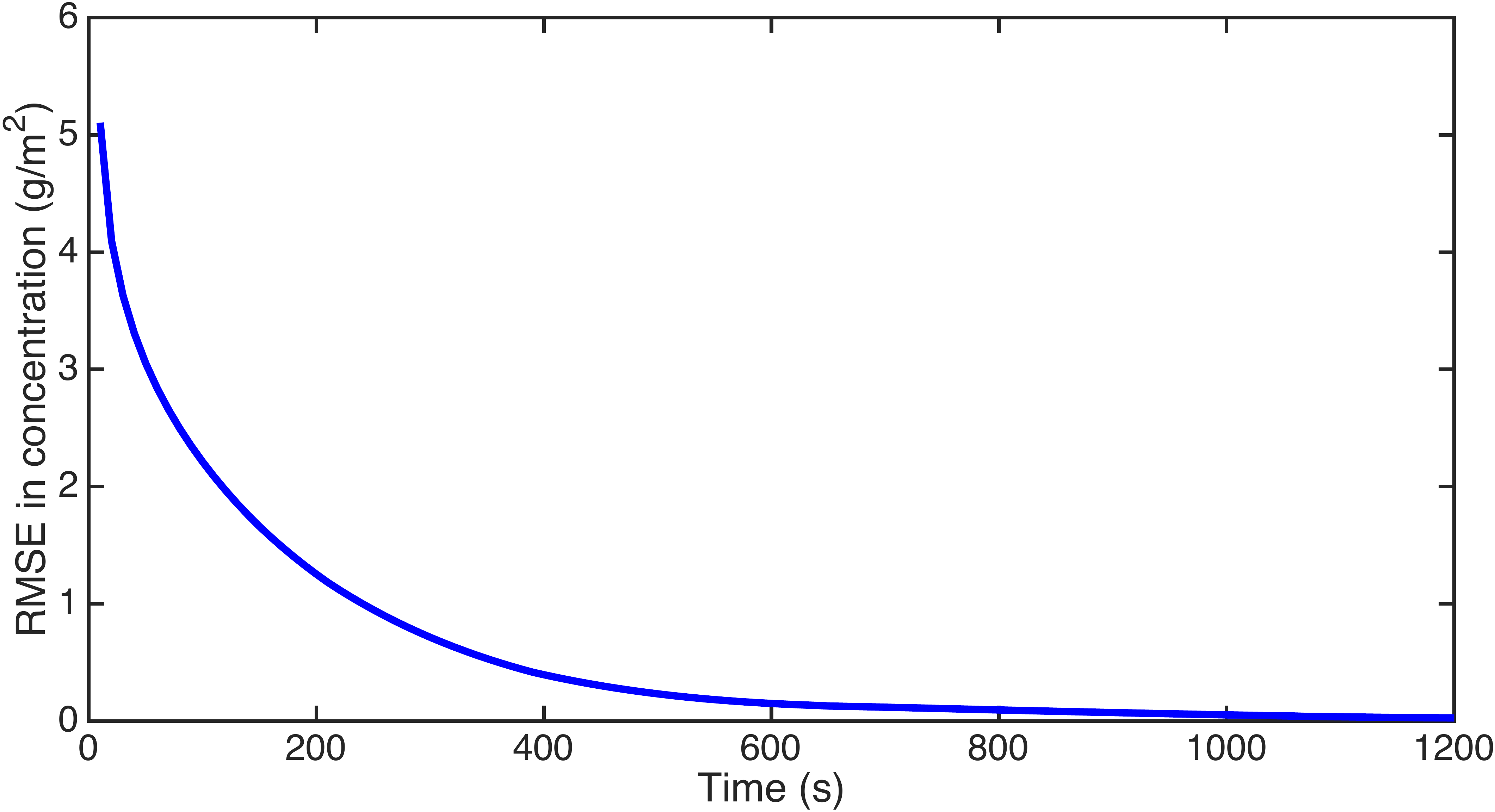}
	\caption{RMSE in concentration of the MAP-MHE state estimator as a function of time, for a random network of 5 threshold sensors.}
	\label{fig:fig1}
\end{figure}

Fig. \ref{fig:fig1} shows the performance of the novel MH-MAP state estimator implemented in MATLAB, in terms of Root Mean Square Error (RMSE) of the estimated concentration field, i.e.:
\begin{equation}\label{64}
\text{RMSE}(t)=\left(\sum_{j=1}^{\alpha}\frac{\|e_{t,j}\|^{2}}{\alpha}\right)^{\frac{1}{2}},
\end{equation}
where $\|e_{t,j}\|$ is the norm of the estimation error at time $t$ in the $j-$th simulation run, averaged over $304$ sampling points (evenly spread within $\Omega$) and $\alpha=100$ independent Monte Carlo realizations. The estimation error is computed at time $t$ on the basis of the estimate $\hat{x}_{t-N|t}$.
\begin{figure}[h!]
	\centering
	\includegraphics[width=\columnwidth]{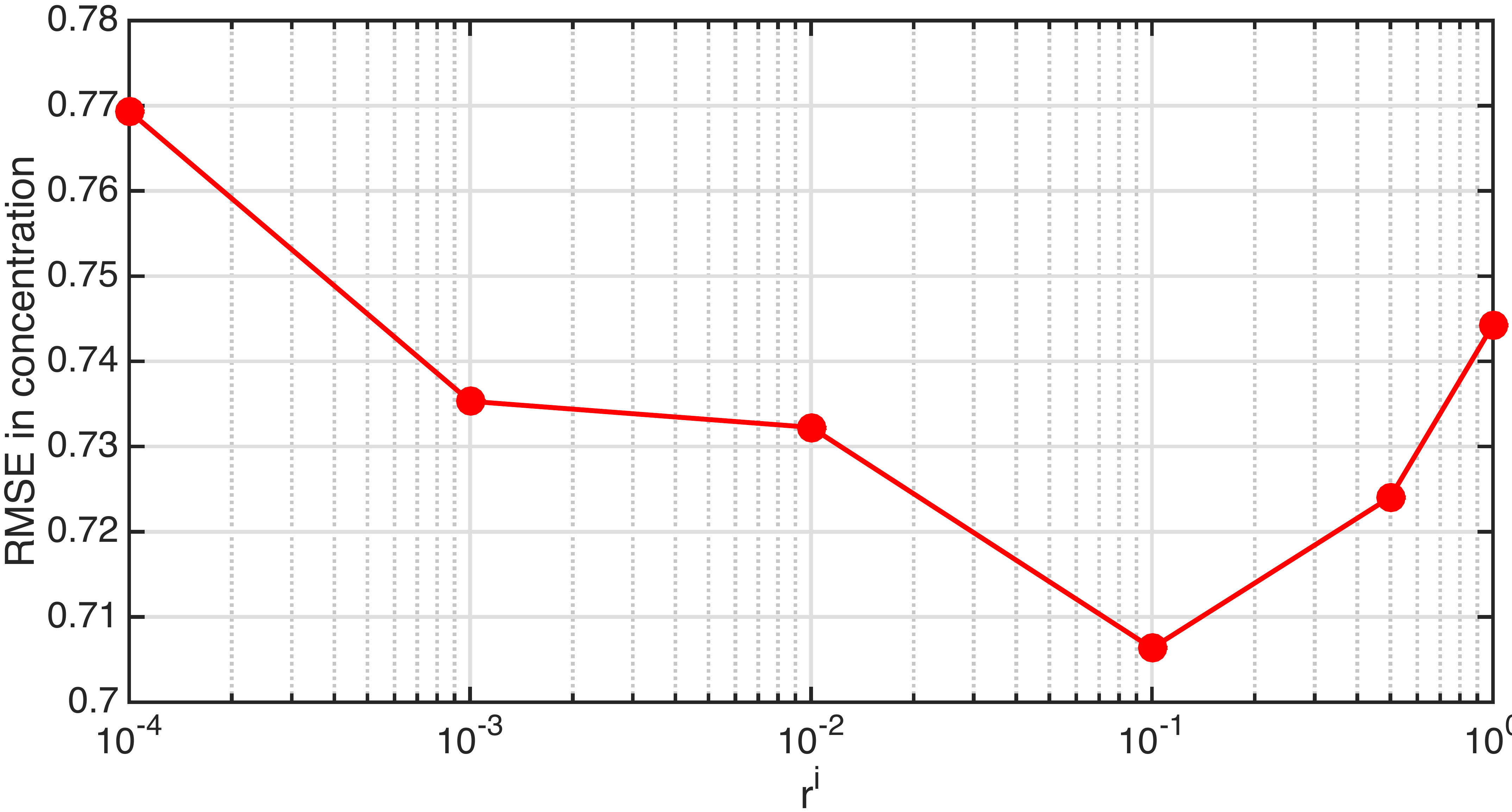}
	\caption{RMSE in concentration as a function of the measurement noise variance, for a fixed constellation of 20 binary sensors. It is shown here that operating in a noisy environment turns out to be beneficial, for certain values of $r^i$, to the state estimation problem.}
	\label{fig:fig2}
\end{figure}
\begin{figure}[h!]
	\centering
	\includegraphics[width=\columnwidth]{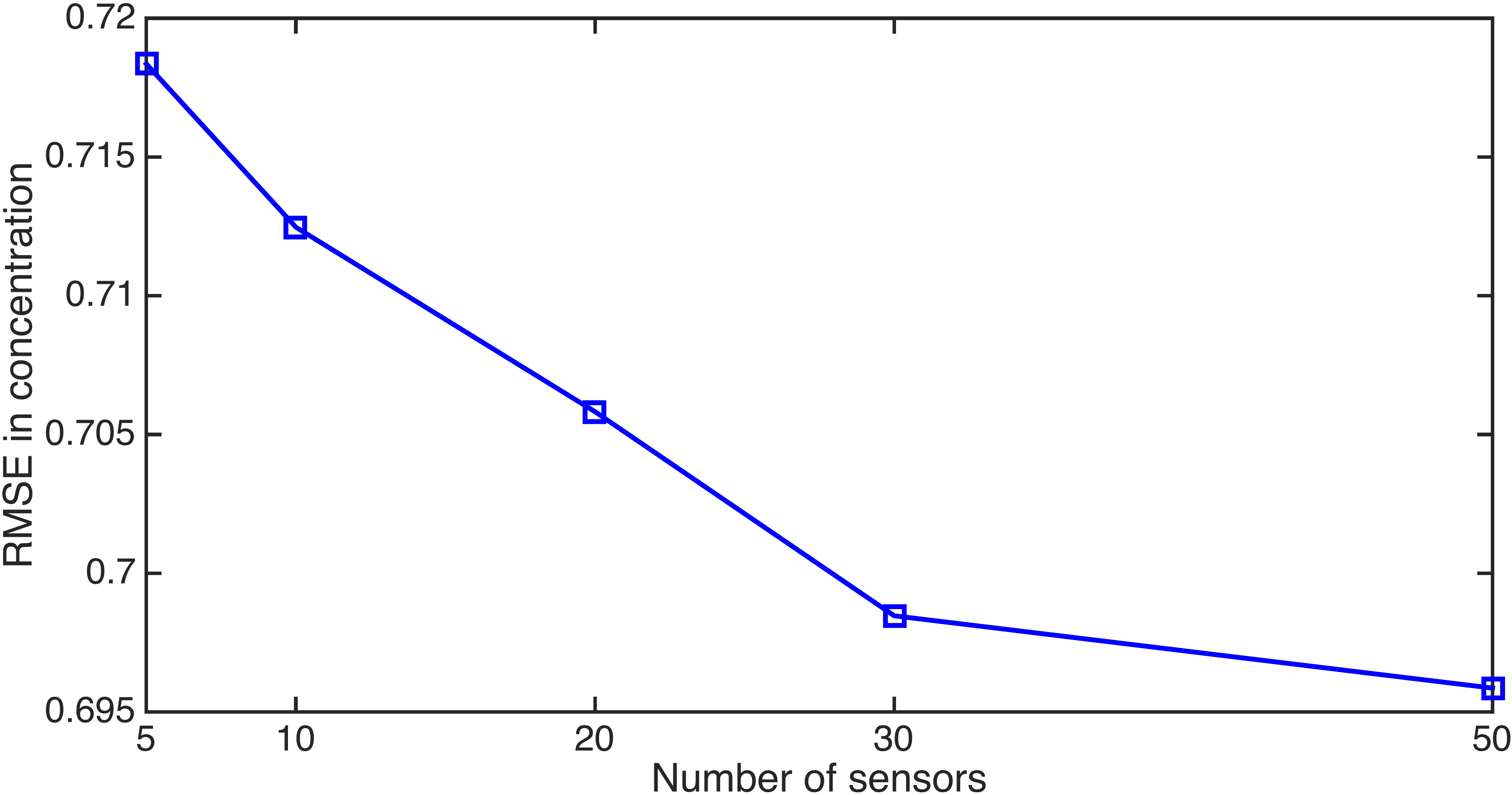}
	\caption{RMSE of the concentration estimates as a function of the number of sensors deployed over the monitoring area.}
	\label{fig:fig3}
\end{figure}
It can be observed that the proposed estimator successfully estimates the dynamic field, even when observed by a network of $l = 5$ randomly deployed binary sensors, with $r^i = 0.25 ~ [g/m^2] ~ \forall i = 1,...,l$.

The effect of measurement noise on the mean value of the RMSE can be seen in Fig. \ref{fig:fig2}, in which it becomes apparent how for certain values of $r^i$, including an observation noise with higher variance, can actually improve the quality of the overall estimates. The results in Fig. \ref{fig:fig2} numerically demonstrates the validity of the above stated noise-aided paradigm in the recursive state estimation with binary measurements and, thus, represents an interesting contribution of this work.
Finally, Fig. \ref{fig:fig3} shows the evolution of the RMSE as a function of the number of binary observations available at the fusion center.

\section{Conclusions and future work}

State estimation with binary sensors has been formulated as a \textit{Moving Horizon} (MH) \textit{Maximum A posteriori Probability} (MAP) optimization problem and it has been shown how such a problem turns out to be convex in the linear system case.
Simulation results relative to a dynamic field estimation case-study have exhibited the conjectured noise-aided feature of the proposed estimator in that the estimation accuracy improves, starting from a null measurement noise, until the variance
of the latter achieves an optimal value beyond which estimation performance deteriorates.

Future work on the topic will concern stability properties of the MH-MAP state estimator and its application to target tracking with binary proximity sensors.

 \addtolength{\textheight}{-7.5cm}   

\section*{Appendix}
\vspace{3mm}\textit{Proof of Proposition 1:}
Under assumption A1, the cost function (\ref{MH}) is convex if and only if $F^i(\tau^{i}-C^{i}x_{t})$ and $\Phi^i(\tau^{i}-C^{i}x_{t})$ are log-concave functions, $\forall i=1,\ldots,p$. A function $f:\mathbb{R}^{n}\rightarrow\mathbb{R}$ is log-concave if $f(x)>0$ for all $x$ in its domain and $ln~f(x)$ is concave  \cite{Boyd}, namely
\begin{equation}\label{eq:logconc}
\nabla^{2}ln~f(x)=\frac{1}{f^{2}(x)}\left[\frac{\partial^{2}f(x)}{\partial x^{2}}f(x)-\left(\frac{\partial f(x)}{\partial x}\right)'\left(\frac{\partial f(x)}{\partial x}\right)\right]<0.
\end{equation}
Let us now consider the CDF $\Phi^i(\tau^{i}-C^{i}x_{t})$ and its complementary function $F^i(\tau^{i}-C^{i}x_{t})$, that are positive functions for all $d^{i}_{t}\triangleq \tau^{i}-C^{i}x_{t}$, $i=1,\ldots,l$. From the fundamental theorem of calculus, namely $\frac{\partial}{\partial x}\left(\int^{a(x)}_{b(x)}f(x)dx\right)=f(a(x))\frac{\partial a(x)}{\partial x}-f(b(x))\frac{\partial b(x)}{\partial x}$ where $a(x)$ and $b(x)$ are arbitrary functions of $x$, the first and the second derivatives of the function $F^i(\tau^{i}-C^{i}x_{t})$ with respect to $x_{t}$ are, respectively, equal to
\begin{equation}
\frac{\partial F^i(\tau^{i}-C^{i}x_{t})}{\partial x_{t}}=\frac{C^{i}}{\sqrt{2\pi r^{i}}}e^{-\frac{(\tau^{i}-C^{i}x_{t})^{2}}{2r^{i}}}
\end{equation}
and
\begin{equation}
\frac{\partial^{2} F^i(\tau^{i}-C^{i}x_{t})}{\partial x_{t}^{2}}=\frac{(C^{i} )' C^{i}}{r^{i}\sqrt{2\pi r^{i}}}(\tau^{i}-C^{i}x_{t})e^{-\frac{(\tau^{i}-C^{i}x_{t})^{2}}{2r^{i}}}.
\end{equation}

If $\tau^{i}-C^{i}x_{t}\leq 0$, then $\frac{\partial^{2} F^i(\tau^{i}-C^{i}x_{t})}{\partial x_{t}^{2}}\leq 0$. Hence $\frac{\partial^{2}F^i}{\partial x^{2}}F^i\leq 0$ and, from
(\ref{eq:logconc}), it follows that
 the Q-function $F^i$ is log-concave.
Conversely, if $\tau^{i}-C^{i}x_{t}>0$, the log-concavity of $F^{i}$ depends on the sign of the term
\begin{eqnarray}
&\frac{\partial^{2}F^{i}}{\partial x^{2}}F^{i}-\left(\frac{\partial F^{i}}{\partial x}\right)'\left(\frac{\partial F^{i}}{\partial x}\right) =&\nonumber \\
&\frac{(C^{i})'C^{i}}{2\pi r^{i}}e^{-\frac{(\tau^{i}-C^{i}x_{t})^{2}}{2r^{i}}}\left[\frac{\tau^{i}-C^{i}x_{t}}{r^{i}}\left(\displaystyle{\int^{\infty}_{\tau^{i}-C^{i}x_{t}}}e^{-\frac{u^{2}}{2r^{i}}}du\right)-e^{-\frac{(\tau^{i}-C^{i}x_{t})^{2}}{2r^{i}}}\right].&\nonumber
\end{eqnarray}
From the convexity properties of the function $f(x)=x^{2}/2$, it can be easily verified for any variable $s,k$ that $s^{2}/2\geq -k^{2}/2+sk$, and hence $e^{-s^{2}/2}\leq e^{-sk+k^{2}/2}$ \cite{Boyd}. Then, if $k>0$, it holds that
\begin{equation*}
\int_{k}^{\infty}e^{-\frac{s^{2}}{2}}ds\leq\int_{k}^{\infty}e^{-sk+\frac{k^{2}}{2}}ds=\frac{e^{-\frac{k^{2}}{2}}}{k}.
\end{equation*}
Since $\tau^{i}-C^{i}x_{t}>0$, with a simple change of variable, it can be stated that
\begin{equation}
\frac{\tau^{i}-C^{i}x_{t}}{r^{i}}\left(\displaystyle{\int^{\infty}_{\tau^{i}-C^{i}x_{t}}}e^{-\frac{u^{2}}{2r^{i}}}du\right)\leq e^{-\frac{(\tau^{i}-C^{i}x_{t})^{2}}{2r^{i}}},
\end{equation}
proving, as a consequence, the log-concavity of the Q-function $F^i(\tau^{i}-C^{i}x_{t})$.

By using the complement rule, the cumulative distribution function can be written as $\Phi^i(\tau^{i}-C^{i}x_{t})=1-F^i(\tau^{i}-C^{i}x_{t})\geq 0$ and $\frac{\partial^{2} \Phi^i(\tau^{i}-C^{i}x_{t})}{\partial x_{t}^{2}}=-\frac{\partial^{2} F^i(\tau^{i}-C^{i}x_{t})}{\partial x_{t}^{2}}$. If $\tau^{i}-C^{i}x_{t}>0$, then $\frac{\partial^{2}\Phi^i}{\partial x^{2}}\Phi^i<0$ such that $\Phi^i$ is log-concave. In the remaining case, i.e. $\tau^{i}-C^{i}x_{t}\leq 0$, noting that $\displaystyle{\Phi^{i}=\frac{1}{\sqrt{2\pi r^{i}}}\int^{\tau^{i}-C^{i}(x_{t})}_{-\infty}e^{-\frac{u^{2}}{2r^{i}}}du=\frac{1}{\sqrt{2\pi r^{i}}}\int_{-(\tau^{i}-C^{i}x_{t})}^{\infty}e^{-\frac{u^{2}}{2r^{i}}}du}$, it can be observed that the sign of the term
\begin{eqnarray}
&\frac{\partial^{2}\Phi^{i}}{\partial x^{2}}\Phi^{i}-\left(\frac{\partial \Phi^{i}}{\partial x}\right)'\left(\frac{\partial \Phi^{i}}{\partial x}\right) = \frac{(C^{i})'C^{i}}{2\pi r^{i}}e^{-\frac{(\tau^{i}-C^{i}x_{t})^{2}}{2r^{i}}} &\nonumber \\
&{} \times \left[\frac{-(\tau^{i}-C^{i}x_{t})}{r^{i}}   \left(\displaystyle{\int^{\infty}_{-(\tau^{i}-C^{i}x_{t})}}e^{-\frac{u^{2}}{2r^{i}}}du\right)-e^{-\frac{(\tau^{i}-C^{i}x_{t})^{2}}{2r^{i}}}\right]&\nonumber
\end{eqnarray}
is negative, thus proving the log-concavity of the CDF $\Phi^{i}(\tau^{i}-C^{i}x_{t})$ and the convexity of the whole cost function. \\
\mbox{   }\hfill $\square$

\end{document}